\documentclass[nofootinbib,superscriptaddress,prd,notitlepage,twocolumn]{revtex4-1}

\usepackage[draft]{showlabels}
\usepackage[utf8]{inputenc}
\usepackage{hyperref}
\usepackage{graphicx}
\usepackage{amsmath,amssymb,amsfonts,amsthm,stmaryrd,mathtools,mathtools}
\usepackage{yfonts}
\usepackage{multirow,bigdelim}
\usepackage{dsfont}
\usepackage{color}
\usepackage{graphicx,color,epstopdf,epsfig,psfrag}
\usepackage[dvipsnames]{xcolor}
\usepackage{ mathrsfs }
\usepackage{footmisc}
\usepackage{enumerate}
\usepackage{tikz}
\usetikzlibrary{calc}
\usetikzlibrary{decorations.pathmorphing}
\usetikzlibrary{shapes.geometric}
\usetikzlibrary{arrows,decorations.markings}
\usepackage{bbold}

\hypersetup{
    colorlinks=true,       
    linkcolor=MidnightBlue,          
    citecolor=BrickRed,        
    filecolor=BrickRed,      
    urlcolor=BrickRed           
}
 
\makeatletter
\newcounter{subsubsubsection}[subsubsection]
\renewcommand\thesubsubsubsection{\thesubsubsection .\@alph\c@subsubsubsection}
\newcommand\subsubsubsection{\@startsection{subsubsubsection}{4}{\z@}%
                                     {-3.25ex\@plus -1ex \@minus -.2ex}%
                                     {1.5ex \@plus .2ex}%
                                     {\centering\normalfont\small\textit}}
\newcommand*\l@subsubsubsection{\@dottedtocline{3}{10.0em}{4.1em}}
\newcommand*{\subsubsubsectionmark}[1]{}
\makeatother

\numberwithin{paragraph}{subsection}

\newcommand{\E}{\mathrm{e}}
\renewcommand{\Im}{{\mathrm{Im}}}
\renewcommand{\Re}{{\mathrm{Re}}}

\newcommand{\R}{{\mathbb R}}

\newcommand{\cA}{{\mathcal A}}

\newcommand{\SU}{\mathrm{SU}}

\newcommand{\SO}{\mathrm{SO}}

\renewcommand{\d}{{\mathrm{d}}}

\newcommand{\be}{\begin{equation}}
\newcommand{\ee}{\end{equation}}
\newcommand{\beq}{\begin{eqnarray}}
\newcommand{\eeq}{\end{eqnarray}}
\newcommand{\bes}{\begin{eqnarray}}
\newcommand{\ees}{\end{eqnarray}}

\newcommand{\mat} [2] {\left ( \begin{array}{#1}#2\end{array} \right ) }

\newcommand{\su}{{\mathfrak{su}}}

\newcommand{\Tr}{{\mathrm{Tr}}}
\newcommand{\f}{\frac}

\def\nn{\nonumber}

\renewcommand{\hat}{\widehat}

\renewcommand{\bar}{\overline}

\newcommand{\ellpl}{\ell_\text{Pl}}

\renewcommand{\cosh}{\mathrm{ch}}

\theoremstyle{definition}

\theoremstyle{remark}

\DeclareMathOperator{\arcosh}{arcosh} 

\newcommand{\aldo}{\color{Green} }

\begin{document}

\title{\LARGE Quasi-local holographic dualities\\ in non-perturbative 3d quantum gravity}

\author{{\bf Bianca Dittrich}}\email{bdittrich@perimeterinstitute.ca}
\affiliation{Perimeter Institute, 31 Caroline St North, Waterloo ON, Canada N2L 2Y5}

\author{{\bf Christophe Goeller}}\email{christophe.goeller@ens-lyon.fr}
\affiliation{Laboratoire de Physique, ENS Lyon, CNRS-UMR 5672, 46 all\'ee d'Italie, Lyon 69007, France}
\affiliation{Perimeter Institute, 31 Caroline St North, Waterloo ON, Canada N2L 2Y5}

\author{{\bf Etera R. Livine}}\email{etera.livine@ens-lyon.fr}
\affiliation{Laboratoire de Physique, ENS Lyon, CNRS-UMR 5672, 46 all\'ee d'Italie, Lyon 69007, France}
\affiliation{Perimeter Institute, 31 Caroline St North, Waterloo ON, Canada N2L 2Y5}

\author{{\bf Aldo Riello}}\email{ariello@perimeterinstitute.ca}
\affiliation{Perimeter Institute, 31 Caroline St North, Waterloo ON, Canada N2L 2Y5}

\date{\today}

\begin{abstract}
We present a line of research aimed at investigating holographic dualities in the context of three dimensional quantum gravity within finite bounded regions.
The bulk quantum geometrodynamics is provided by the Ponzano-Regge state-sum model, which defines 3d quantum gravity as a discrete topological quantum field theory (TQFT).
This formulation provides an explicit and detailed definition of the quantum boundary states, which allows a rich correspondence between quantum boundary conditions and boundary theories, thereby leading to holographic dualities between 3d quantum gravity and 2d statistical models as used in condensed matter. 
After presenting the general framework, we focus on the concrete example of the coherent twisted torus boundary, which allows for a direct comparison with other approaches to 3d/2d holography at asymptotic infinity.
We conclude with the most interesting questions to pursue in this framework. 
\end{abstract}

\maketitle

\section{Overview}

In this short paper, we present a research program aimed at understanding the bulk-boundary relationship from the bulk non-perturbative quantum gravity perspective. 
Our goal is to clarify the role of boundaries and  the physics of edge modes in quantum gravity. 
In non-perturbative approaches to quantum gravity, it is natural to consider finite, or `quasi-local' boundaries. 
On these, we can derive the boundary theory induced by the choice of classical boundary conditions or class of quantum boundary states.
In a holographic approach, these boundary theories define holographic duals, still encoding the same physical content as the bulk theory, providing a non-trivial representation of bulk observables and allowing to reconstruct the bulk quantum geometry from boundary correlations. 
Our objective is to test this non-perturbative holographic approach, which provides a quasi-local version of the AdS/CFT correspondence \cite{Witten:1998qj}.

Three-dimensional gravity is an ideal testbed for this program. 
Indeed, it can be formulated as a topological field theory, which allows for an exact non-perturbative quantization \cite{Witten:1988hc}. 
In this context, we propose to study the Ponzano--Regge state-sum model \cite{PonzanoRegge1968,Freidel:2004vi,Freidel:2005bb,Barrett:2008wh}.
This is a three-dimensional discrete topological quantum field theory (TQFT), whose relation to the Turaev--Viro \cite{Turaev:1992hq} and Reshetikhin--Turaev topological invariants is explicitly understood \cite{reshetikhin1990ribbon,ReshetikhinTuraev1991,Freidel:2004nb}. 

Concretely, the Ponzano--Regge model provides a quantized path integral for 3d Regge calculus, which is a well-defined discretization of general relativity \cite{Regge1961,Regge:2000wu}.
Specifically, it corresponds to a theory of quantum gravity with vanishing cosmological constant in Euclidean signature. Generalizations of the model to non-vanishing cosmological constant exists \cite{Turaev:1992hq,MizoguchiTada1992,TaylorWoodward2005,Freidel:2005bb}, but we postpone their investigation to future work.

From the Ponzano--Regge state-sum, transition amplitudes for physical states of quantum geometry can be explicitly computed.
However, the question of holographic dualities has not yet been systematically explored in this framework, despite being a very natural one.
This is because, the Ponzano--Regge model provides a bulk-local description of the 3d quantum geometry. It is thus natural to focus on finite bounded regions of space-time, explore the various possible quantum boundary conditions and investigate the resulting quantum boundary theories.

Since the Ponzano-Regge model is intrinsically discrete
, we expect the Ponzano-Regge amplitude with boundaries to induce discrete statistical physics models on the 2d boundary. Such models typically lead to 2d conformal field theories (CFTs) in their continuum limit in their critical regime. This scenario would lead to a quasi-local version of the AdS${}_{3}$/CFT${}_{2}$ correspondence \cite{brown1986central,Carlip2005b,Heydeman:2016ldy,Sfondrini:2014via,Kraus:2006wn}.

An important question is whether the continuum limit  can be reached for finite boundaries at all. 
Indeed, quantum gravity naturally sets a fundamental minimal length scale and for this reason it is likely that the infinite refinement limit needed for reaching the continuum can not be achieved for finite boundaries.
In turn, this would mean that the dual statistical physics boundary theories can only reach their critical regime for asymptotic boundaries. 
In other words, they would flow towards their limit CFT as the typical length scale of the boundary geometry is much larger than the fundamental quantum gravity length scale, and to the extent that the latter scale can be neglected. 
For finite boundaries, we would then be left with the non-critical discrete statistical models as holographic duals.
Their correlations should still allow to probe the bulk 3d quantum geometry and faithfully represent the 3d quantum gravity amplitudes.

At this point a remark is in order. Three-dimensional gravity is topological, i.e. it does not possess any local physical degrees of freedom. For this reason, discrete models such as the Ponzano--Regge topological state-sum can capture all the relevant degrees of freedom of the continuum theory, irrespectively of the coarseness of the employed discretization.
%
%
However, the introduction of a boundary with metric boundary conditions does reveal the discretization, which explains our remarks in the previous paragraph.\\

In this paper, we  present and streamline our recent results derived in details \cite{PRholo1,PRholo2} and push their analysis further.  In an effort to clarify the context of that work, we explain how to derive the holographic duals on the boundary of the Ponzano--Regge model (see also \cite{Riello2018}), and discuss the critical problem of identifying a quantum notion of asymptotic infinity. In particular, we discuss how we are able to reproduce results found in perturbative contexts \cite{Barnich:2015mui, BonzomDittrich} while at the same time extending these results by non-perturbative corrections.

{
Since in the Ponzano--Regge model there are two ways of increasing the size of the boundary and the number of degrees of freedom it carries, two naive notions of large scale limit exist: one consists in taking a large number of Planck-sized building blocks, the other in taking a large value for the spins $j$ at fixed number of building blocks. 
In the former approach, a renormalization flow can be defined that describes effective actions taking the effects of a finer and finer lattice into account (see e.g. \cite{Dittrich:2014ala}).
On the top of this refinement limit, a semi-classical limit can also be taken.
The latter approach, instead, turns out to be best understood not in in terms of a large-scale limit, but rather in terms of a semiclassical (e.g. 0 or 1-loop) limit on a fixed discretization (e.g. \cite{PonzanoRegge1968,Bianchi:2006uf,Livine:2006ab,Han:2013hna,Barrett:2009mw}).
On the top of this semiclassical limit, a continuum limit can also be taken. 
Therefore, what seemed to be two different ways of taking a large scale limit, are rather two different limits altogether.
\footnote{
For completeness, let us mention an extra possibility with regards to the continuum limit, which we will not pursue here. It is possible to consider the discretization itself as a quantum degree of freedom to be ``summed over''. This is actually the main insight for the dynamical triangulation approach to quantum gravity. In the context of spinfoam models, this point of view leads to group field theory and tensor models, for which one define and study a renormalization flow \cite{Carrozza:2013mna,Rivasseau:2011hm}. It is an open question how these various limits and renormalization flows connect to each other. 
}
When both are taken one after the other, the result will not a priori be independent of the order in which the limits are taken.
 Nonetheless, we showed in \cite{PRholo1,PRholo2} that some important features agree in the two approaches.
Indeed, both these choices remarkably lead to the same divergence behavior of the associated amplitudes  \cite{PRholo1,PRholo2}. Crucially, the same behavior that characterizes the perturbative results as well. In the case of the semi-classical calculation, the one-loop result not only reproduces the perturbative ones obtained both in the continuum \cite{Barnich:2015mui} and in the discrete \cite{BonzomDittrich}, but also includes contributions of saddle points associated to non-classical `quantum' backgrounds.

We hope that the present program will also help elucidate renormalization in discrete quantum gravity models \cite{Dittrich:2014ala, Dittrich:2014mxa,Riello:2013bzw,Banburski:2014cwa} and in particular provide first steps to implement holographic renormalization \cite{deBoer:1999tgo,deHaro:2000vlm} in non-perturbative quantum gravity models.
}

\section{Statistical Duals}

The Ponzano--Regge model \cite{PonzanoRegge1968,Freidel:2004vi,Freidel:2005bb,Barrett:2008wh} proposes a topological path integral for discretized 3d geometries.
Initially defined for triangulations, it can readily be extended to arbitrary 3d cellular decomposition. The amplitudes are well-defined through suitable gauge-fixing \cite{Freidel:2004vi,Bonzom:2010zh,Bonzom:2012mb} and define a topological invariant \cite{Barrett:2008wh,Freidel:2004nb}. They have two equivalent definitions, either in terms of products of spin recoupling symbols (such as the $\{6j\}$-symbol representing a quantized tetrahedron) or as a discretized path integral over holonomies encoding the parallel transport along the triangulated manifold. Here, we will not review these definitions of the  Ponzano--Regge bulk amplitude, but we will focus on the boundary states and induced boundary theory.

The general relation between boundary conditions, or boundary states, and (dual) boundary theories, is to be looked for in the correspondence between spin-network states and statistical models {\aldo \cite{Witten1989,Witten1990, Turaev1992, KirillovReshetikhin1989, Westbury1998, Dittrich:2013jxa, Bonzom:2015ova, Riello2018}}.
 
Spin-network states are gauge-invariant Wilson graph observables naturally associated to the connection representation of gravity.
Denoting by $\omega$ the relevant connection 1-form and by $g_{l_i} = P\exp\int_{l_i} \omega$ the parallel transport along the $i$-th link $l_i$ of the Wilson graph $\Gamma$, such states have the form
\be
\Psi[\omega] = \Psi^\Gamma(\{g_{l_i}\}_i) = \Psi^\Gamma(\{ h_{t(l_i)} g_{l_i} h_{s(l_i)}^{-1}\}_i ),
\ee
where the last equality is a representation of gauge invariance, with $s(l)$ and $t(l)$ denoting the source and target vertices of $l$, respectively.
Using standard loop quantum gravity techniques, when the supporting graph is dual to a surface, spin-network states---or some specific superpositions thereof---can be interpreted as quantum boundary metrics. 

Being supported on a graph, these are intrinsically discrete objects.
It is important to notice, however, that the Hilbert space of spin-network states in loop quantum gravity can be understood as a space of {\it continuum} quantum connections,\footnotemark~ and in this sense all spin-network states are continuum states.
Nonetheless, their {\it operational} geometrical  interpretation still reposes on discrete structures, and their discreteness can be interpreted as the result of a ``physical'' finite-resolution pre- or post-selecting measurement.
In this case, the graph itself can be interpreted quite literally as a network of physical beacons and space-measuring devices.%
\footnotetext{This is guaranteed by cylindrical consistency of the wave functions together with the existence of an inductive limit on the wave functions: spin networks are actually defined on infinite sequences of refining graphs. The inductive continuum limit can moreover be defined both for the Ashtekar--Lewandowski and $BF$ representations of LQG \cite{Ashtekar:1994mh,Baez:1994hx,Freidel:2011ue,Bahr:2015bra}.}

On top of the graph-induced discreteness, when dealing with Euclidean theories, another type of discreteness comes into play.
This is the Planck-scale discreteness associated to the {\it spectrum} of metric operators.
The mathematical origin of this discreteness lies in the compactness of the parallel transport variables $g_l$ along the edges of the graph, which replaced their infinitesimal counterparts---the connection variable $\omega$---as the fundamental variable. 

In three Euclidean dimensions, $g_l\in\SU(2)$ and the specifics of the quantum metric described by a spin-network state are encoded in the choice of ($i$) $\SU(2)$ spins $j_l$ attached to the edges of the spin-network graph, and of ($ii$) gauge-invariant tensors, aka intertwiners $\iota_v$, attached to its vertices.
Spin-network states with fixed spins and intertwiners will be denoted by
\be
\Psi^\Gamma_{(j,\iota)}(g_l).
\ee
More specifically, intertwiners are $\SU(2)$-invariant states in the tensor product of several $\SU(2)$ representations, say labeled by $N$ spins $j_{1},..,j_{N}$,
\be
\iota_v=  \bigotimes_{i=1}^N D^{j_i}(h) \triangleright \iota_v
\,,\quad\forall h\in\SU(2)
\,,
\ee
that is, explicitly showing the sum over magnetic indices,
\be
(\iota_v)_{n_1\dots n_N} = \sum_{\{m\}}\Big( \prod_{i=1}^N D^{j_i}(h)_{n_i m_i}\Big) (\iota_v)_{m_1\dots m_N}
\,.
\ee
With this notation, $\Psi^\Gamma_{(j,\iota)}(g_l)$ is defined as the contraction of the vertex intertwiners with the $j_l$-representation Wigner matrices of the parallel transports $g_{l}$, according to the combinatorics imposed by the underlying graph $\Gamma$:
\be
\Psi^\Gamma_{(j,\iota)}(g_l)
\,=\,
\mathrm{Tr}_\Gamma \Big(\bigotimes_{v}\iota_{v}\otimes \bigotimes_{l}D^{j_{l}}(g_{l})\Big)
\,,
\ee
(here the trace stands for the sum over the magnetic indices at both ends of each link of the graph $\Gamma$.)

The amplitude $Z_M$ of a quantum gravitational process in a finite region $M$ subjected to the boundary conditions imposed by a given spin-network $\Psi^\Gamma$, is of course a function of the spin-network state itself. 
When dealing with the dynamics of flat space, that is of three-dimensional (quantum) gravity with vanishing cosmological constant, this function is extremely simple.
If $M=\mathbb B_3$ with the 2-sphere boundary, it essentially reduces to what is known as a spin-network evaluation, i.e.
\be
Z_{\mathbb B_3}(\Psi^\Gamma) = \Psi^\Gamma(g_l = \mathbb{1}).
\ee
Crucially, this evaluation gets twisted in non-trivial ways by the presence of topological features such as (bulk) non-contractible cycles \cite{Freidel:2005bb}. 

Spin-network evaluations are complete contractions of intertwiners, associated to the vertices of the spin-network graph. 
Such an evaluation can be read as the sum over all magnetic-index configurations one can attach to the edges of the graph,  with specific Boltzmann weights given by the entries of the intertwiner tensors themselves. Symbolically,
\be
Z_{\mathbb B_3}(\Psi^\Gamma_{(j,\iota)}) = \mathrm{Tr}_\Gamma \Big( \bigotimes_{v\in\Gamma} \iota_v \Big) = \sum_{\{m_l | l\in \Gamma\}} \prod_{v\in\Gamma} (\iota_v)_{m\dots} 
\label{eq_sum_m}
\ee
where in the last term the $m$-indices of $\iota_v$ label states in those representations $V_{j_l}$ of $\SU(2)$ which are attached to the links of $\Gamma$ starting at or leaving from the vertex $v$.
From this perspective, $\SU(2)$ spin-network evaluations is essentially the computation of a partition function of a statistical model, whose rotational invariance is automatically guaranteed by the $\SU(2)$-invariance of the intertwiners themselves.

\begin{figure}[h!]
	\begin{tikzpicture}[scale=1]
	\draw[decoration={markings,mark=at position 0.3 with {\arrow[scale=1.5,>=stealth]{>}}},decoration={markings,mark=at position 0.8 with {\arrow[scale=1.5,>=stealth]{>}}},postaction={decorate}]  (0,-1) node[left ]{\tiny{2}}-- (0,1)node[right ]{\tiny{1}}; \draw[decoration={markings,mark=at position 0.3 with {\arrow[scale=1.5,>=stealth]{>}}},decoration={markings,mark=at position 0.8 with {\arrow[scale=1.5,>=stealth]{>}}},postaction={decorate}] (-1,0)node[above ]{\tiny{3}}--(1,0)node[below ]{\tiny{4}};
	\draw (0,-1.5) node[scale=1]{$\omega(\mathrm I) =a$};
	
	\draw[decoration={markings,mark=at position 0.3 with {\arrow[scale=1.5,>=stealth]{<}}},decoration={markings,mark=at position 0.8 with {\arrow[scale=1.5,>=stealth]{<}}},postaction={decorate}]  (0,-4) -- (0,-2); 
	\draw[decoration={markings,mark=at position 0.3 with {\arrow[scale=1.5,>=stealth]{<}}},decoration={markings,mark=at position 0.8 with {\arrow[scale=1.5,>=stealth]{<}}},postaction={decorate}] (-1,-3)-- (1,-3);
	\draw (0,-4.5) node[scale=1]{$\omega(\mathrm{II}) =a$};

	\draw[decoration={markings,mark=at position 0.3 with {\arrow[scale=1.5,>=stealth]{>}}},decoration={markings,mark=at position 0.8 with {\arrow[scale=1.5,>=stealth]{>}}},postaction={decorate}]  (3,-1) -- (3,1); 
	\draw[decoration={markings,mark=at position 0.3 with {\arrow[scale=1.5,>=stealth]{<}}},decoration={markings,mark=at position 0.8 with {\arrow[scale=1.5,>=stealth]{<}}},postaction={decorate}] (2,0)-- (4,0);
	\draw (3,-1.5) node[scale=1]{$\omega(\mathrm{III}) =b$}; 
	
	\draw[decoration={markings,mark=at position 0.3 with {\arrow[scale=1.5,>=stealth]{<}}},decoration={markings,mark=at position 0.8 with {\arrow[scale=1.5,>=stealth]{<}}},postaction={decorate}]  (3,-4) -- (3,-2); 
	\draw[decoration={markings,mark=at position 0.3 with {\arrow[scale=1.5,>=stealth]{>}}},decoration={markings,mark=at position 0.8 with {\arrow[scale=1.5,>=stealth]{>}}},postaction={decorate}] (2,-3)-- (4,-3);
	\draw (3,-4.5) node[scale=1]{$\omega(\mathrm{IV}) =b$};
	
	\draw[decoration={markings,mark=at position 0.3 with {\arrow[scale=1.5,>=stealth]{<}}},decoration={markings,mark=at position 0.8 with {\arrow[scale=1.5,>=stealth]{>}}},postaction={decorate}]  (6,-1) -- (6,1); 
	\draw[decoration={markings,mark=at position 0.3 with {\arrow[scale=1.5,>=stealth]{>}}},decoration={markings,mark=at position 0.8 with {\arrow[scale=1.5,>=stealth]{<}}},postaction={decorate}] (5,0)-- (7,0);
	\draw (6,-1.5) node[scale=1]{$\omega(\mathrm{V}) =c$};
	
	\draw[decoration={markings,mark=at position 0.3 with {\arrow[scale=1.5,>=stealth]{>}}},decoration={markings,mark=at position 0.8 with {\arrow[scale=1.5,>=stealth]{<}}},postaction={decorate}]  (6,-4) -- (6,-2); 
	\draw[decoration={markings,mark=at position 0.3 with {\arrow[scale=1.5,>=stealth]{<}}},decoration={markings,mark=at position 0.8 with {\arrow[scale=1.5,>=stealth]{>}}},postaction={decorate}] (5,-3)-- (7,-3);
	\draw (6,-4.5) node[scale=1]{$\omega(\mathrm{VI}) =c$};
	
	\end{tikzpicture}
	\caption{The 6 possible arrow configurations at a vertex for the 6-vertex model. The statistical weights $\omega$ associated to the vertex configurations are $\omega(I)= \omega(II) = a$, $\omega(III)= \omega(IV) = b$ and $\omega(V)= \omega(VI) = c$. For the correspondence with the spin network evaluation defined the Ponzano-Regge partition function for an homogeneous spin $j=\f12$ on the boundary lattice, the arrow direction translates into the sign of the magnetic index $m$ living on the edge. Indeed, for a spin $j=\f12$, the magnetic index can only take two values, $m=\pm \f12$.}
	\label{fig:6vertex_model_vertex}
\end{figure}
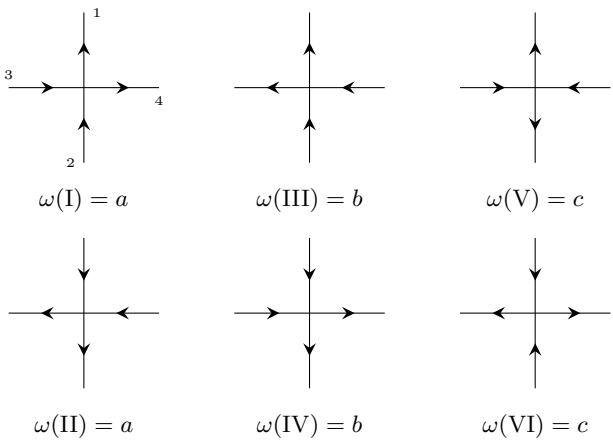

In particular, for homogeneous spins $j=1/2$ for all edges on the boundary spin network, when the graph is a regular square lattice, which we denote $\Gamma=\square$, we can show the correspondence of the Ponzano-Regge amplitude with boundary with the partition function of the ``isotropic'' 6-vertex model \cite{PRholo1} (see also \cite{Witten1989,Witten1990} for a broader perspective and equivalence of 3d quantum gravity with statistical models), as illustrated on figure \ref{fig:6vertex_model_vertex}:
\be
Z(\Psi^\square_{(\f12,\iota)})
= \sum_\text{arrows} a^{\#_I + \#_{II}} b^{\#_{III}+\#_{IV}} c^{\#_{V} + \#_{VI}}
\ee
with
\be
\Delta := \frac{a^2 + b^2 - c^2}{2ab} =1.
\ee
In the case of non-trivial topologies,  the above equality still holds locally, but further non-local operator insertions are needed to account for the existence of non-trivial holonomies along the non-contractible cycles. We explore this case in the next section (see also \cite{Riello2018}).

This is an integrable statistical model, whose transfer matrix coincides with that of the XXX Heisenberg spin-chain with spectral parameter $\lambda = i(a/c - 1/2)$, e.g. \cite{PinkBook}.
The degrees of freedom of the spin-chain are indeed the same magnetic indices appearing in Eq. \eqref{eq_sum_m}, however the system is not quite in a pure Gibbs ensamble of the form $\Tr(e^{-\beta H_\text{XXX}})$, but rather in some involved generalized ensamble, whose ``Hamiltonian'' is given by a combination of conserved quantities weighted by $\lambda$, e.g. \cite{Faddeev}.
Notice that the XXX spin chain is isotropic and hence $\SO(3)$ invariant.
See \cite{Riello2018} for more on these correspondences.

For generic spins, a special class of spin-network graphs is constituted by Wilson line weaves.
In a weave, a set of Wilson lines pass above and beneath each other, forming links of knots, without ever intersecting. 
This situation corresponds to intertwiners whose recoupling channel is in a state of vanishing spin (notice that these form a basis of intertwiners among four spins $j=1/2$).
Such peculiar states were observed to be related to integrable models of various kinds almost thirty years ago \cite{KirillovReshetikhin1989,Turaev1992, Witten1989,Witten1990}.
Nevertheless, at the time, the perspective was different, and the correspondence was made with knot expectation values in Chern--Simons theory, rather than with boundary spin-network evaluations.%
\footnote{
The degrees of freedom in those statistical models are more naturally expressed in terms of the spins $j$, rather than the magnetic indices $m$. This `spin representation' has also a holographic/gravitational interpretation, as explained in \cite{Riello2018}.
}
%

The appearance of magnetic indices as degrees of freedom of the boundary theory is an aspect that deserves attention.
Magnetic indices are quantum states in a co-adjoint orbit of $\mathfrak{su}^\ast(2)$. 
Therefore, following the LQG quantum geometrical interpretation, they represent the possible (quantum) orientations of a vector of fixed length
in $\mathbb R^3$.
Thus, the present correspondence explicitly fulfils the expectation of much of the contemporary work in the context of gravity and gauge theories in presence of boundaries, that the dual boundary degrees of freedom are constituted by reference frame orientation of a ``would-be-gauge'' symmetry
\cite{DonnellyFreidel2016, GomesRiello2017, Geiller2017b,Carlip2005a,Carlip2005b, Balachandran1992,Balachandran1996}.


For increasing values of the spins, the number of allowed magnetic indices grows: the number of Planck-sized cells present in the larger co-adjoint orbits grows, and the latter can be better and better approximated by their classical counterparts. 
This explains why, in the large spin regime, semiclassical methods exist to analyze the relevant spin-network evaluations.
These methods are indeed well developed, both in the mathematical and physical literature, and the emergence of geometrical objects from the large spin asymptotics of such evaluations is well studied \cite{PonzanoRegge1968, Roberts1999,Gukov:2003na,TaylorWoodward2005,Barrett:2009gg,Barrett:2009mw,Haggard:2014xoa,Barrett:1998gs}.

So far, we considered only spin-network states at fixed spins.
This, however, need not be a necessary restriction, and considering superpositions of all spins has its own interest.
For example, this fact can be used to peak the state not only on an intrinsic geometry, corresponding to metric boundary conditions, but also on its extrinsic geometry, leading to more general boundary conditions \cite{BahrThiemann2007, Bianchi:2006uf}.

Another interesting example is given by the construction of so-called spin-network generating functions.
These are boundary states which depend analytically on one parameter per edge, in such a way that their power expansion in these parameters gives all possible (fixed-spin) spin-network evaluations.
Pictorially, these parameters can be interpreted as chemical potentials---or, more precisely, fugacities---for the edge spins.
Crucially, generating-function boundary states on specific graph types encode the (bosonized dual of the) Ising model \cite{Bonzom:2015ova, Dittrich:2013jxa}.
%

\section{Non-trivial Topologies}

As it was mentioned earlier, these correspondences are enriched by the presence of boundary non-contractible cycles, which stay non-contractible in the bulk.
At this point a clarification is necessary.

It is often assumed that in a theory of quantum gravity, one has to sum over {\it all} manifolds compatible with the boundary data, including all compatible bulk topologies.
However natural and appealing, this idea is plagued with difficulties ranging from defining evolution on non-globally-hyperbolic manifolds in a canonical setting, to the very classification of topologies in dimensions 4 and higher in a covariant setting.
For these reasons, in a large part of the quantum gravitational literature (see however \cite{Rivasseau:2011hm,Carrozza:2013mna, MaloneyWitten2007}), the issue of summing over topologies has been put aside, and the focus restricted to the ``integration'' over metrics.%
\footnote{In four dimensions, the problem is even subtler, because of the non-uniqueness and proliferation of differential structures.}
Here, we will adopt this conservative viewpoint, and hence postpone all investigations of topology change, and large diffeomorphisms alike.

This being clarified, we can address the question of how the boundary theory ``knows'' about its bulk-contractible and bulk-non-contractible cycles. 
From the bulk perspective, non-contractible cycles correspond to non-trivial monodromies that need to be integrated over.
Using gauge invariance, these monodromies can be made to have support on a single small cylinder $\mathbb D_2\times [0,\epsilon]$, whose boundary counterpart is a ring $\mathbb S^1\times[0,\epsilon]$ winding around the conjugate boundary cycle.
This ring supports a non-local ``topological'' operator acting on the boundary spin-network state. 
The topological nature of this operator is a consequence of the flatness of the fundamental connection.
More technically, at the spin-network level, the operator arising from the integration over all possible monodromies---when unconstrained---can be easily recognized to correspond to the insertion along the ring of a Haar intertwiner or, in other words, of a group-averaging operator%
\footnote{A Haar intertwiner is an intertwiner of the form $$(\iota_\text{Haar})_{m'_1\dots m'_L}^{m_1\dots m_L}=\int_{\SU(2)} \d g\, \prod_{l=1}^L D^{j_l}(g)^{m_l}{}_{m'_l},$$ where $D^j(g)$ is a Wigner matrix in the representation of spin $j$; $m,m'\in\{-j,\dots, j\}$ are magnetic indices; $\d g$ is the Haar measure on the group. Finally, in this formula, $l$ labels the edges crossing the above-mentioned ring.\label{fnt_D}}%
---see \cite{PRholo1,PRholo2,Riello2018}.

The modification of the boundary theory via this operator insertions explicitly breaks the symmetry between the various non-contractible cycles of the surface, thus imprinting on the boundary theory some knowledge of the bulk topology.  
In practical computations, these topological operators play a crucial role.
We will present a concrete example of this in the next section.

\section{Example: Coherent Torus}

The simplest example that can provide insights on the above program is given by a twisted solid torus spacetime in three-dimensional Euclidean quantum gravity with vanishing cosmological constant. Twisted means that the solid torus $M_3 \cong \mathbb D_2\times\mathbb{S}_1$ is obtained by identifying the bottom and the top of the cylinder $\mathbb D_2(a)\times [0,\beta]$ up to a rotation of $\gamma$ radians. Here, $a$ stands for the radius of the two-disk, and $\beta$ for the Euclidean time extension of the cylinder.

This example has already been studied with other techniques---most notably via the perturbative quantum Einstein--Hilbert theory at 1-loop over a flat background, both in the continuum \cite{Barnich:2015mui} and in the ``discretum'' \cite{BonzomDittrich}, and as a limit of the corresponding AdS$_3$/CFT$_2$ computation \cite{MaloneyWitten2007,GiombiMaloneyYin2008,BarnichOblak2014,Oblak:2015sea}---and can therefore serve as a benchmark for the present methods.

The most prominent feature of all these approaches, is a partition function which depends on $\gamma$ in a way that admits a (formal) expansion over boundary momentum eigenmodes $p\in\mathbb N$---i.e. Fourier modes in the bulk-contractible, ``spacelike'', direction---as%
\footnote{The identification of $p$ as spacelike eigenmodes is clearest in some formulations \cite{MaloneyWitten2007,Oblak:2015sea,BonzomDittrich}, but can remain rather obscure in others \cite{GiombiMaloneyYin2008,Barnich:2015mui}. }
\be
Z(\beta,\gamma) \sim \E^{-\frac{(2)\pi \beta}{\ellpl}} \prod_{p\geq2}\frac{1}{2 -2 \cos(\gamma p)} ,
\label{eq_Z}
\ee
where $\ellpl=8\pi G_\text{N}\hbar$, and the factor of 2 is in parenthesis because its presence depends on the chosen boundary conditions (standard Gibbons--Hawking--York boundary conditions require it; for details, see the first section of \cite{PRholo1}).

Notice the peculiar fact that for $\gamma \in 2\pi \mathbb Q$, there are $p$'s whose contribution explodes.

In AdS space, this issue---as well as the convergence of $Z$ at $p\to\infty$---is cured by the fact that the role of $\gamma$ is played by the torus' modulus, $2\pi\tau = \gamma + i\sqrt{|\Lambda|}\beta$.
The resulting ``thermal AdS'' partition function is then closely related to Dedekind's $\eta(\tau)$ function, a modular form strictly speaking defined  on the upper half complex plane, $\Im(\tau)>0$ \cite{MaloneyWitten2007}. In this respect the flat limit is expected to be quite singular, and deserves to be studied independently.

In any case, quite remarkably, even in the flat case multiple different approaches \cite{Barnich:2015mui,BonzomDittrich,Oblak:2015sea} agree with the formal result of Eq. \eqref{eq_Z}, although the mechanisms leading to this  formula and the regularizations that make sense of it are technically very different. 
In all cases, a crucial fact is that the modes $p=0$ and $p=1$ do not appear in the product as a consequence of diffeomorphism symmetry (see \cite{PRholo1} for a detailed comparison).  

Maybe one of the most interesting derivations of the formula \eqref{eq_Z} is as a character of BMS$_3$ in the ``vacuum'' representation (``massive'' ones have a different prefactor, and a product over $p$ that starts at $p=1$) \cite{Oblak:2015sea}.
This is taken as a hint that a dual boundary theory indeed exists whose symmetry group is (a central extension of) BMS$_3$, in analogy to the Virasoro symmetry of the AdS$_3$/CFT$_2$ case.
This idea is further supported by the fact that the BMS$_3$ characters can be obtained through a zero-cosmological-constant limit of the Virasoro ones.


Here, we present results on the analysis of precisely this situation within the bulk local non-perturbative approach provided by the Ponzano--Regge (PR) model \cite{PRholo1,PRholo2}.

The first question one faces in setting up the computation within the PR model is what boundary state one is going to use.
A survey of the previous method suggests that the boundary state should encode the geometry of a rectangular cylinder.
Remarkably, this fact is made explicit in the only other quasi-local computation \cite{BonzomDittrich}, while it is implicitly used in the other computations intrinsic to the boundary---since they assume translational symmetry along the two boundary directions. It is, however, completely hidden in the perturbative Einstein--Hilbert approach.

In order to impose the desired boundary conditions, coherent spin network techniques, developed in the context of loop quantum gravity were used to design a state corresponding to an  $N_t\times N_x$ regular rectangular lattice.
%
%
The twisted toroidal topology  is implemented by an identification of the lattice appropriately shifted by $N_\gamma$ units in the ``spatial'' direction (see figure \ref{fig:discretization_exemple}), so that
\be
\gamma = 2\pi \frac{ N_\gamma}{N_x},
\ee
while the spin-network intertwiner encoding the rectangular plaquette geometry dual to the spin-network vertex $v$ is\footnote{See footnote \ref{fnt_D} for details on the $D$-matrix notation. Notice that the second magnetic index, $m'_l$ in the notation of footnote \ref{fnt_D}, is here fixed to its maximal value $j_l$. This choice ultimately ensures that these states are coherent states.}
\be
\iota_v^{m_1\dots m_4} = \int_{\SU(2)}\d G_v \prod_{l=1}^4  D^{j_l}(G_v g_{\xi^v_l})^{m_l}{}_{j_l}.
\label{eq_coh}
\ee
\begin{figure}[h!]
	\begin{tikzpicture}[scale=0.55]
	\coordinate (OA) at (1.59,0);
	\coordinate (A1) at (0,0);
	\coordinate (A2) at (1.1,0.77);
	\coordinate (A3) at (2.7,0.77);
	\coordinate (A4) at (3.3,0);
	\coordinate (A5) at (2.1,-0.83);
	\coordinate (A6) at (0.5,-0.83);
	
	\coordinate (OB) at (1.59,-1.5);
	\coordinate (B1) at (0,-1.5);
	\coordinate (B2) at (1.1,-0.73);
	\coordinate (B3) at (2.7,-0.73);
	\coordinate (B4) at (3.3,-1.5);
	\coordinate (B5) at (2.1,-2.35);
	\coordinate (B6) at (0.5,-2.35);
	
	\coordinate (OC) at (1.59,-3);
	\coordinate (C1) at (0,-3);
	\coordinate (C2) at (1.1,-2.23);
	\coordinate (C3) at (2.7,-2.23);
	\coordinate (C4) at (3.3,-3);
	\coordinate (C5) at (2.1,-3.83);
	\coordinate (C6) at (0.5,-3.83);
	
	\draw (A1) -- (A2) -- (A3) -- (A4) -- (A5) -- (A6) -- cycle; 
	\draw (OA) -- (A1) ; \draw (OA) -- (A2); \draw (OA) -- (A3); \draw (OA)--(A4); \draw (OA) --(A5); \draw (OA)-- (A6); 
	
	\draw (B1) -- (B2) -- (B3) -- (B4) -- (B5) -- (B6) -- cycle;
	\draw [dashed] (OB) -- (B1); \draw [dashed] (OB) -- (B2); \draw [dashed] (OB) -- (B3); \draw [dashed] (OB)--(B4); \draw [dashed] (OB) --(B5); \draw [dashed] (OB)-- (B6);
	
	\draw (C1) -- (C2) -- (C3) -- (C4) -- (C5) -- (C6) -- cycle;
	\draw [dashed] (OC) -- (C1); \draw [dashed] (OC) -- (C2); \draw [dashed] (OC) -- (C3); \draw [dashed] (OC)--(C4); \draw [dashed] (OC) --(C5); \draw [dashed] (OC)-- (C6);
	
	\draw (A1)--(B1)--(C1); \draw [dashed] (A2)--(B2)--(C2); \draw [dashed] (A3)--(B3)--(C3); \draw (A4)--(B4)--(C4); \draw (A5)--(B5)--(C5); \draw (A6)--(B6)--(C6);
	\draw [dashed] (OA)--(OB)--(OC); 
	
	\draw [<->,>=latex] (-0.5,0) -- (-0.5,-3); \draw (-0.5,-1.5) node[scale=0.8,left]{$\beta$};
	\draw [<->,>=latex] (0,1.5) -- (1.59,1.5); \draw (0.8,1.5) node[scale=0.8,above]{$a$};

	\draw (1.5,-5) node{$(a)$}; 
	
	\end{tikzpicture}
\hspace*{10mm}
	\begin{tikzpicture}[scale=0.4]

	
	\draw (4.5,0) -- (13,0);  
	\draw (4.5,-1.5) -- (13,-1.5);
	\draw (4.5,-3) -- (13,-3);
	
	\draw (5,1) -- (5,-4);
	\draw (6.5,1) -- (6.5,-4);
	\draw (8,1) -- (8,-4);
	\draw (9.5,1) -- (9.5,-4);
	\draw (11,1) -- (11,-4);
	\draw (12.5,1) -- (12.5,-4);
	
	\draw (5.75,0) node[above]{$T$}; \draw (5,-0.75) node[right]{$L$};
	
	\foreach \i in {5,6.5,8,9.5,11}{
		\draw[rounded corners=3 pt,->] (\i,2) --(\i,1.7)-- (\i+1.5,1.6)--(\i+1.5,1.1)   ;
	}
	
	\foreach \i in {5,6.5,8,9.5,11,12.5}{
		\foreach \j in {0,-1.5,-3}{
			\draw (\i,\j) node{$\bullet$};	
		}
	}
	\draw (5,-4.5) node{$1$}; \draw (8,-4.5) node{...}; \draw (12.5,-4.5) node{$N_x = 6$};

\draw (8.7,-6.3) node{$(b)$};
	
	\end{tikzpicture}
	\caption{Discretization of the twisted solid torus for parameters $N_t=3$, $N_x=6$ and a shift $N_\gamma=1$.	
		$(a)$: On the left hand side, we draw the cylinder with base the 2-disk of radius $a$ and with Euclidean time extension $\beta$. To get the twisted solid torus, we identify the top and the bottom of the cylinder up to  the discrete shift $i\rightarrow i+N_{\gamma}$ for all lattice sites $i$, which causes the twist $\gamma = 2\pi \f{N_\gamma}{N_x}$ in the gluing.
		$(b)$: On the right hand side, we draw the boundary lattice associated to the  discretization of the solid torus. We attach a spin $T$ (resp. $L$) to each horizontal (resp. vertical) edges on the boundary, and we attach to each boundary vertex $v$ an intertwiner as defined by formula \eqref{eq_coh}. }
\end{figure}
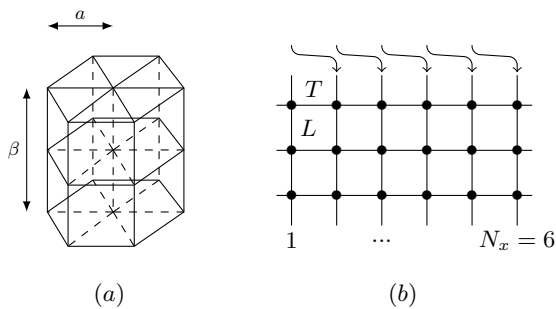

Here, $j_l = T\in\tfrac12 \mathbb N$ ($ L\in\tfrac12 \mathbb N$) for horizontal ``$\text{h}$'' (vertical ``$\text{v}$'') edges $l$ dual to ``timelike'' (``spacelike'') sides of the rectangular plaquettes, and the normalized spinors $\xi^v_l\in \mathbb C^2$ are
\be
\mat{c}{1\\0}, \;\frac{1}{\sqrt 2}\mat{c}{-1\\1},\; \mat{c}{0\\1}, \text{ or } \frac{1}{\sqrt 2} \mat{c}{1\\1},
\ee
for $l$ ranging from $1$ to $4$ in an anti-clockwise order around the spin-network vertex, starting from an horizontal edge. 
Finally,
\be
 g_\xi = \mat{cc}{\xi^1 &  -\bar\xi^2 \\ \xi^2 & \bar \xi^1}\in\SU(2),
\ee
where the bar stands for complex conjugation.

Eq. \eqref{eq_coh} defines a coherent intertwiner, designed to encode the geometry of a flat rectangle.
The four spinors $\xi_\ell$ represent in a precise sense the sides of the rectangular plaquette along the $\hat z,\, -\hat x, \, -\hat z,$ and $\hat x$ directions respectively, while the spins $L,\,T$ correspond to their lengths, and the lift of the group elements $G_v\in\SU(2)$ to $\mathrm{SO}(3)$ corresponds to the orientation of the plaquette's reference frame.%
\footnote{Using the fact that at each vertex $\sum_{l=1}^4 j_l$ is an integer, one can show that $G_v\mapsto (-1) G_v $ is a symmetry of the integrand. Using this fact, one can consistently consider the $G_v$ to be elements of $\SU(2)/\mathbb{Z}_2\cong \SO(3)$, hence truly representing frame orientations. This identification will be implicitly assumed in the following treatment.}
Gauge invariance implies that all orientations are weighed equally.
Being coherent means that in the large spin regime $T,L\to\infty$ uniformly, the above state optimally minimizes the extrinsic curvature along the rectangle diagonals compatibly with Heisenberg uncertainty principle---see e.g. \cite{KapovichMillson1996,barbieri1998quantum}.

The PR amplitude for such a boundary state $\Phi_\text{coh}$ can be written, after some manipulation and gauge fixing, as the following purely boundary theory
\be
Z_\text{PR}(\Phi_\text{coh})=\int \d \varphi \left[\prod_{v} \int \d G_{v}\right] \sin^2\left(\frac\varphi2\right)\,\E^{\sum_{l} S_l }
\,,
\label{eq_Zcoh}
\ee
where the label ``coh'' stands for ``coherent''. The angle $\varphi\in[0,2\pi[$ corresponds to the conjugacy class of the only non-trivial holonomy wrapping around the non contractible cycle of the cylinder.

Furthermore,  $S_l$ denotes the contribution to the ``action'' of the spin-network edge $l$:
\be
S_l \hspace{-2pt}=\hspace{-3pt}
\begin{cases}
2 j_l \ln \langle \xi^{t(l)}_l| G^{-1}_{t(l)}G_{s(l)} |\xi^{s(l)_l}\rangle &\hspace{-7pt} \text{$l$ horiz.}\\
2 j_l \ln \langle \xi^{t(l)}_l| G^{-1}_{t(l)}\E^{-i\frac{\varphi}{2N_t} \sigma_z}G_{s(l)} |\xi^{s(l)_l}\rangle & \hspace{-7pt}\text{$l$ vert.}
\end{cases}
\ee
Here, the edges $l$ are oriented to point either to the right or upwards, and $v=s(l), \,t(l)$ are therefore its source and target vertices, respectively. 
Notice that the branch-cut of the logarithm plays no role.

The $G_v$ are $\SU(2)$ elements associated to the vertices $v$ of the boundary coming from \eqref{eq_coh}; in this formulation, they constitute the degrees of freedom of the dual field theory, replacing the magnetic indices of the statistical model formulation (the two formulations are in the end equivalent, see  \cite{Riello2018} for a more thorough discussion of this point). Geometrically, the $G_v$ can be understood as providing a ``potential'' for a flat boundary connection and thus describe the embedding of the boundary into flat 3d space.

Remark now that the action $\sum_l S_l$ is complex and such that $\Re(S_\text{coh})\leq0$.
One can think of its imaginary part as providing the actual action for the boundary theory, and its real part as providing the quantum measure.\footnote{The action is imaginary although the signature of the gravitational field is Euclidean. This is because the Ponzano--Regge model is a quantization of three dimensional gravity in the form of a topological $\SU(2)$ $BF$-theory. In this formulation, the physical signature of the metric is purely encoded in the gauge group.
}
For notational simplicity, the two contributions will not be explicitly distinguished.

Now, using that $j_l$ is the eigenvalue
\footnote{Other operator orderings can be used, for instance in the standard loop quantum gravity spectrum is given by the square root of the Casimir, $\ell_{l} = \ellpl \sqrt{j_l(j_l+1)}$.} 
in Planck units of the length operator along (the dual of) $l$:
\be
\ell_{l} = \ellpl\, j_l,
\ee
one sees that in the semiclassical limit $\hbar \to 0$, at fixed boundary geometry---that is at fixed $\ell_{l}$---the coherent intertwiners get peaked on the classical geometry, while the amplitude $Z_\text{PR}$ can be evaluated at 1-loop via a stationary phase approximation.

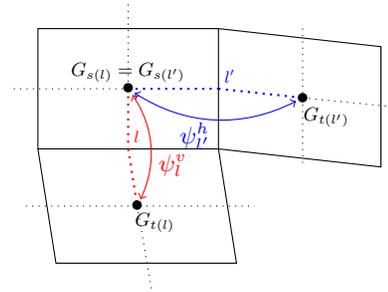
\begin{figure}[h!]
	\begin{center}
		\begin{tikzpicture}[scale=0.80,line/.style={<->,shorten >=0.1cm,shorten <=0.1cm}]
		\coordinate(a1) at (0,0); \coordinate(b1) at (0.3,-1.9);
		\coordinate(a2) at (3,0); \coordinate(b2) at (3.3,-1.9); \coordinate(d1) at (5.7,-0.3);
		\coordinate(h1) at (0,2); 
		\coordinate(h2) at (3,2); \coordinate(d2) at (5.7,1.7);

		\draw (a1)--(a2)--(h2)--(h1)--cycle;
		\draw (a1)--(b1)--(b2)--(a2);
		\draw (a2)--(d1)--(d2)--(h2);
		
		\draw[dotted] (1.5,2.4)--(1.5,1);
		
		\draw[dotted] (-0.4,1)--(1.5,1); 
		
		\draw[dotted,blue,thick] (1.5,1)--(3,1); \draw[blue] (3,1) node[scale=0.7,above right]{$l'$}; \draw[dotted, blue, thick] (3,1) -- (4.4,0.85);
		\draw[dotted,thick,red] (1.5,1)--(1.5,0); \draw[red] (1.5,0) node[scale=0.7,above right]{$l$}; \draw[dotted,red,thick] (1.5,0)--(1.65,-0.95);

		\draw[dotted] (1.65,-0.95)--(1.9,-2.4);
		\draw[dotted] (-0.2,-0.95)--(3.6,-0.95);
		
		\draw[dotted] (4.4,0.85)--(5.9,0.7); \draw[dotted] (4.4,2)--(4.4,-0.3);
		
		\draw (1.5,1) node[above, scale=0.8]{$G_{s(l)} = G_{s(l')}$}; \draw (1.5,1) node{$\bullet$};
		\draw (1.5,-0.95) node[below right,scale=0.8]{$G_{t(l)}$};\draw (1.65,-0.95) node{$\bullet$};
		\draw (4.3,0.8) node[below right,scale=0.8]{$G_{t(l')}$};\draw (4.4,0.85) node{$\bullet$};
		
		\path [red,line,bend left] (1.5,1) edge node[midway,below right]{$\psi_{l}^{v}$} (1.65,-0.95);
		\path [blue,line,bend right] (1.5,1) edge node[midway,below left=-0.7mm]{$\psi_{l'}^{h}$} (4.4,0.85);
		\end{tikzpicture}
	\end{center}
	\caption{Three plaquettes - faces - of the boundary discretization dual to 3 neighboring vertices. The dotted line are the links of the dual lattice. In red, the dihedral angle $\psi_{l}$ along the  link $l$ relates the group elements $G_{s(l)}$ and $G_{t(l)}$ living on the two corresponding plaquettes, which are vertical neighbors. In blue, the dihedral angle $\psi_{l'}^{h}$ along the horizontal link  $l'$ relates the group elements $G_{s(l')}$ and $G_{t(l')}$ . }
	\label{fig:psi_as_dihedral_angle}
\end{figure}

The stationary phase equations for the boundary action are $\Re(\sum_l S_l)=\text{max}_{\{G_v\}}\left(\Re(\sum_l S_l)\right)=0$, and $\delta_{G_v}\left(\sum_l S_l\right)=0$ which is satisfied if and only if
\be
\begin{cases}
G_{s(l)} \xi_l^{s(l)}= \E^{\frac{i}{2} \psi^\text{h}_l} G_{t(l)} \xi_l^{t(l)} & \text{$l$ horizontal}\\
G_{s(l)} \xi_l^{s(l)}= \E^{\frac{i}{2} \psi^\text{v}_l} \E^{-\frac{i}{2}\frac{\varphi}{N_t} \sigma_z}G_{t(l)} \xi_l^{t(l)} & \text{$l$ vertical}.
\end{cases}
\ee
Using standard spinfoam techniques \cite{Barrett:1998gs,Freidel:2002mj,Livine:2006it,Livine:2007vk,Barrett:2009mw}, these equations can be geometrically interpreted as describing an immersion (i.e. a local embedding) of the tiled toroidal surface in the ambient flat space $\mathbb R^3$ \cite{Dowdall:2009eg,2011arXiv1103.5644C}.
According to this interpretation, $\hat G_v =\mathrm{Ad}_{G_v}$ represents the $\mathrm{SO}(3)$ frame of the rectangular plaquette dual to $v$---which is defined up to some global rotation along the $\hat z$ axis%
\footnote{The restriction to the $\hat z$ axis is a consequence of a gauge fixing. Generally this is the axis picked by the non-trivial bulk holonomy.}%
---while $\psi_l$ represents the dihedral angle (extrinsic curvature) between two neighboring plaquettes connected by $l$ (see figure \ref{fig:psi_as_dihedral_angle}).

The equation resulting from the stationarity of the bulk holonomy's conjugacy class $\varphi$, $\delta_\varphi\left(\sum_l S_l\right)=0$ i.e.
\be
\hat z\cdot \sum_{v} \hat G_v\triangleright \hat x= 0 ,
\ee
breaks the symmetry between the two cycles of the torus (here implicit in the appearance of $\hat x$ rather than $\hat y$ in the equation), and implies that 
\be
\psi^\text{v}_l = 0 \quad\text{if $l$ vertical}.
\ee
(An alternative solution is $\psi_l=\pi$, always for $l$ vertical; solutions where this happens are called ``folded'' solutions \cite{PRholo2} and will be ignored in this article---however, cf. the comment at the end of this section on Planck-scale values of the extrinsic curvature).

The above equation means that the equation of motion for $\varphi$ dictates along which cycle the torus boundary can be curved extrinsically, and this happens precisely in the spatial direction, as it was intuitively expected.

Thus, the solutions to the equations of motion are given by rectangular cylinders whose ``spatial'' sections are (not-necessarily convex) $N_x$-sided polygons and whose ends are identified modulo a shift of $N_\gamma$ units.

Boundary conditions further require that along each ``time slice''
\be
\sum_{x=1}^{N_x} \psi^\text{h}_{l_x} = 0 \;\text{mod}\;2\pi
\ee
and
\be
\varphi = \sum_{x=1}^{N_\gamma} \psi^\text{h}_{l_x} \;\text{mod}\;2\pi
\ee
where the latter equation is understood to hold for {\it any} choice of $N_\gamma$ consecutive horizontal edges $\{l_x\}$.

Fixing $N_x$ and $N_\gamma$ to be coprime,
\be
K\equiv\mathrm{GCD}(N_x,N_\gamma)=1,
\ee
the analysis of the stationary phase equations can be pushed further. 
In particular, $K=1$ implies that the polygonal sections must be regular, since it implies that
\be
\psi_l^\text{h} = \frac{2\pi}{N_x}n \;\;\; \text{for all horizontal $l$'s}
\ee
for some $n\in\{0,\pm1,\dots,\pm\tfrac{N_x-1}{2}\}$, and hence
\be
\varphi = 2\pi \frac{N_\gamma}{N_x} n = \gamma n.
\ee

For $K\neq 1$, all solutions are part of a continuous family, and the 1-loop determinant (the Hessian determinant of $\sum_l S_l$ at the stationary point) contains null directions. Geometrically, these degeneracies correspond to the fact that one can deform the polygonal section of the cylinder while staying on-shell.

Back to the case $K=1$, the integer $n$ labels the different stationary points. 
We interpret it geometrically  as a winding number which counts how many times the toroidal surface  winds around the cylinder's axis before closing. 
The existence of these solutions is due to the fact that the models relies on (compactified) holonomy rather than connection variables, and consequently---loosely speaking---the flatness condition just states that the deficit angle must be an integer multiple of $2\pi$, rather than strictly zero.
The degenerate case $n=0$ is suppressed because of the integration measure, while the case $n=1$ is of course the ``geometrical'' one. 

At these stationary points, the spin-network action takes the on-shell value (recall that for vertical links, $\psi^\text{v}_l=0$)
\be
\sum_l S_l =   i N_t \sum_{x=1}^{N_x} T \psi^\text{h}_{l_x} = 2\pi i n T N_t.
\ee
Formally, this action has the structure of a discretized Gibbons--Hawking--York term on the rectangular cylinder, as desired from what should correspond to the on-shell value of the Einstein--Hilbert action on a flat spacetime and in agreement with Eq. \eqref{eq_Z}. However, once exponentiated, $\E^{\sum_l S_l}$ reduces to a sign. This is a consequence of the discreteness of the spin $T\in\frac12\mathbb N$.

If $K=1$, moreover, the 1-loop determinant is non-degenerate and can be readily analyzed.
Using a Fourier transform adapted to the presence of the twist, one can diagonalize the Hessian, while an exact resummation formula gets rid of the explicit energy dependence. 
The ensuing result for the determinant, combined with the on-shell value of the $\varphi$ measure, is
\be
(\text{1-loop det}) \times \sin^2\frac\varphi2 = \mathcal A(n)\times \mathcal D(\gamma,n),
\ee
where the $\gamma$ dependence is fully contained in 
\be
\mathcal D(\gamma, n) = (2-2\cos\gamma n ) \prod_{p=1}^{\frac{N_x-1}{2}}\frac{1}{2-2\cos\gamma p}.
\label{eqnDp}
\ee
When focusing on the geometrical background geometry corresponding to $n=1$, this reproduces precisely the sought result of Eq. \eqref{eq_Z}.
For this to work, it is crucial to notice the fundamental role played by the integration over the non-trivial holonomy wrapping around the non-trivial cycle of the torus, which contributes precisely the factor $2-2\cos \gamma n$ above.

Interestingly, the product over $p$ in the equation above can be explicitly computed once we remember that the angle $\gamma$ comes from the shift $N_{\gamma}$ in the gluing of the lattice:
\be
\prod_{p=1}^{\frac{N_x-1}{2}}\frac{1}{2-2\cos\,\f{2\pi N_{\gamma}p}{N_{x}} }
=
N_{x}
\,
\ee
(this formula holds as long as $N_{\gamma}$ and $N_{x}$ are coprime, i.e. $K=1$). This gives a surprisingly simple result for the amplitude for a given winding number $n$,
\be
\mathcal D(\gamma, n) = 2N_{x}\,(1-\cos\gamma n) \,.
\label{eqnD}
\ee
Notice that, despite this simplification, the dependence on the twist $\gamma$ does not completely disappear and we keep a non-trivial result.

A first remark is that it is very interesting that our lattice computation leads to such a straightforward finite truncation of the BMS$_3$ character formula for the partition function \eqref{eq_Z} to the product over modes $p$ bounded by the lattice size $N_{x}$. Moreover the simplification of this product is a great coincidence, which likely points out towards a underlying powerful symmetry. This point needs to be investigated further.

A second remark is that the equation \eqref{eqnDp} is actually more useful than the simplified expression  \eqref{eqnD} in order to understand the physical content of the theory. Indeed, it provides the mode decomposition of the theory, in terms of the Fourier modes $p$. Although the partition function $Z$ might simplify, what matters is that the various Fourier modes have different weights with a specific $\gamma$-dependence, i.e. $(2-2\cos\gamma p)^{-1}$, which is probed e.g. by the correlations of the boundary theory.

In this sense and to this extent, this calculation perfectly agrees with the $\Lambda\to 0$ limit of the AdS case given  by eq. \eqref{eq_Z} when focusing on the geometrical background corresponding to the classical solution with winding number  $n=1$.

Finally, the amplitude factor ${\mathcal A}(n)$ carries the dependence on the winding number $n$. It is a rather intricate function of the spins $L$ and $T$, the lattice sizes $N_{x}$ and $N_{t}$ and of course of the label $n$. It is nevertheless possible to considerably simplify the formula derived in \cite{PRholo2} (by explicitly performing the product over Fourier modes $p$). The result is
\beq
\cA(n)
&=&
\left[\f{iL - (L+T)\tan\f{\psi n}{2} }{i L - (L +4TN_tN_{x}) \tan\f{\psi n}{2}}\right]^{\f12}
\\
&&\times
\,
\left[
\f{2(2\pi)^{3}e^{in\psi}}{LT(L+T)}
\right]^{\f{N_{x}N_{t}}2}
\Big[2T_{N_{x}}(a_{n})-2\Big]^{-\f{N_{t}}2}
\,,
\nn
\eeq
where $\psi$ is the  dihedral angle unit for the lattice and  $a_{n}$ is a simple complex trigonometric function,
\be
\psi=\f{2\pi}{N_{x}}
\,,\quad
a_{n}=\cos n\psi+\f{iL}{T+L} \sin n\psi
\,.
\ee
Also, the notation $T_{N}(a)$ stands for the Chebyshev polynomial (of the first kind) of order $N$.%
\footnote{
Recall, $T_{N}(a)=\cosh (N\arcosh a)$ and\\ 
$\Big(2T_{N}(a)-2\Big)^{\f12}
=
\left(a+\sqrt{a^{2}-1}\right)^{\f N2}
-\left(a+\sqrt{a^{2}-1}\right)^{-\f N2}
\,.$
}

From these expressions one can first check the reality property
\be
\mathcal A(-n) = \overline{\mathcal A}(n)
\,.
\ee
This is consistent with a Hamilton--Jacobi functional, which cannot distinguish a momentum from its opposite, and more specifically with a first-order Einstein--Cartan formulation of General Relativity, of which the Ponzano-Regge model is a quantization. 

\begin{figure}[t]
	\begin{center}
		\includegraphics[height=5cm]{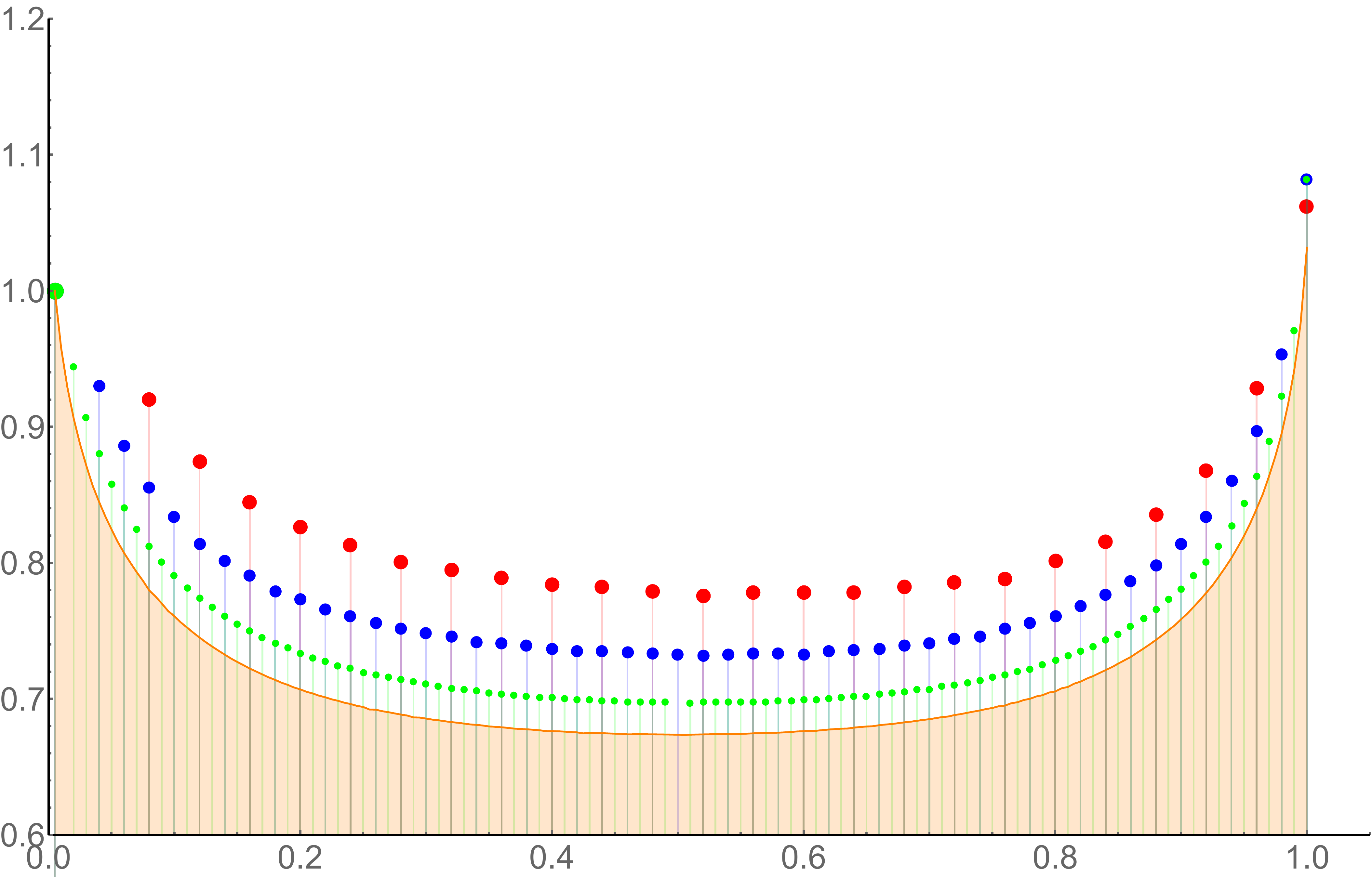}
	\end{center}
	\caption{Plots of $\frac{\log(|\mathcal A(1)|)}{\log(|\mathcal A(n)|)}$ for $n$ running from $1$ to $\frac{N_x-1}{2}$ with the parameters $N_t=20$, $L=8$, $T=8$. The four plots correspond to $N_x=50,100,200,400$ (red, blue, green, orange). The $x$-axis corresponds to $\frac{2n}{N_x-1}$, which runs from 0 to 1.}
	\label{fig:ALS_plot}
\end{figure}

\begin{figure}[h!]
	\begin{center}
		\includegraphics[height=4cm]{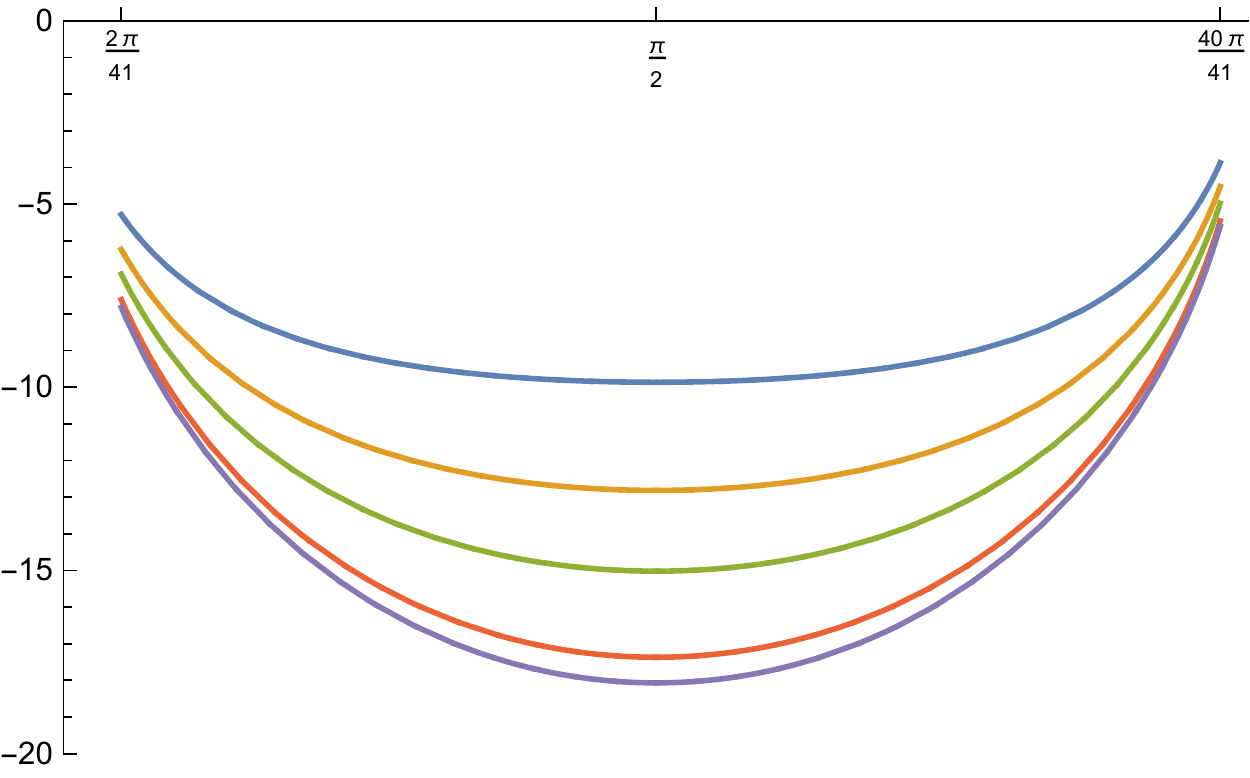}
	\end{center}
	\caption{
	Plot of ${\f{-N_t}2}\log\big{|}2T_{N_{x}}(a_{n})-2)\big{|}$ in terms of the continuous angle variable $x=n\psi\in\left[\f{2\pi}{N_x},\pi-\f{\pi}{N_x}\right]\subset[0,\pi]$ for the odd lattice size $N_{x}=41$ and $N_{t}=1$ and for spins $T=5$ and $L=5,10,20,100$. As $L$ increases and thus the ratio $\f TL$ goes to 0, the curves gets more and more curved and goes to the limit function (lowest curve).
	}
	\label{fig:chebplot}
\end{figure}

\begin{figure}[h!]
	\begin{center}
		\includegraphics[height=25mm]{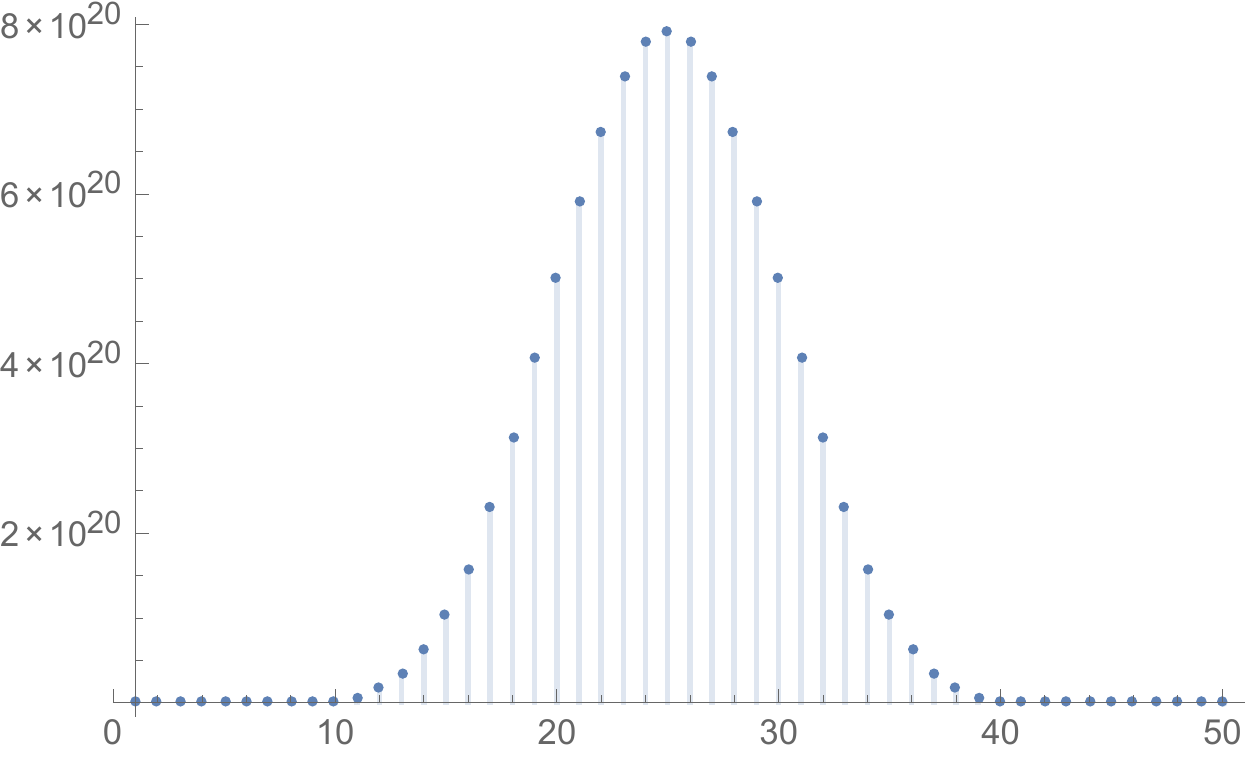}
		\hspace*{2mm}
		\includegraphics[height=25mm]{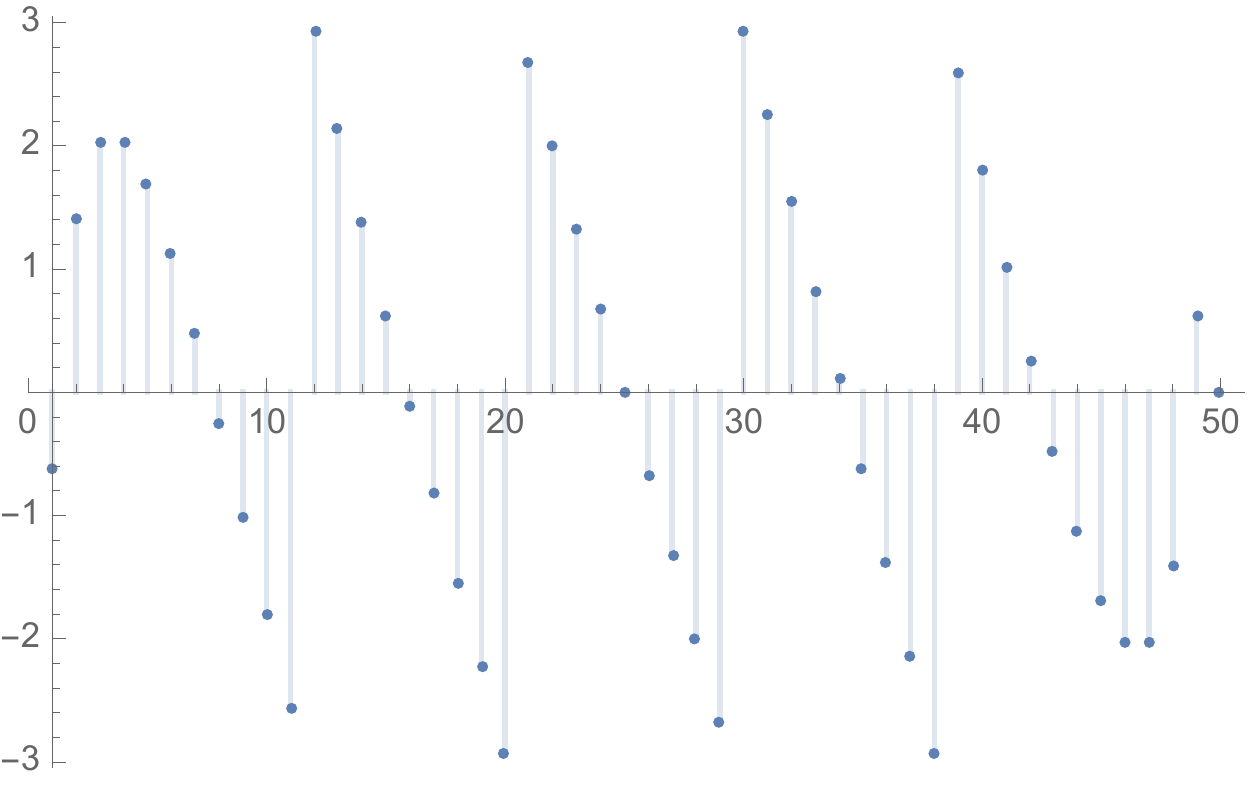}
	\end{center}
	\caption{Plots of the modulus (on the left) and argument (on the right) of $\big{(}2T_{N_{x}}(a_{n})-2)\big{)}$ for $N_{x}=100$ with $n$ running from 1 to 50, for spin parameters $L=T=1$.}
	\label{fig:chebmodargplot}
\end{figure}

Moreover, ${\cal A}(n)$ is, in the large $N_x, N_t$ limit (but already true when they are larger than 5),  overwhelmingly peaked in modulus at the minimal and maximal values of $n$, that is at $n=1$ and $n=\lfloor\frac{N_x-1}{2}\rfloor$
as illustrated by the plots on  figure \ref{fig:ALS_plot}.
This behavior is entirely due to the factor with the Chebyshev polynomial, $\big{(}2T_{N_{x}}(a_{n})-2)\big{)}^{-\f{N_{t}}2}$. 
We can actually plot this at fixed $N_{x},N_{t}$ as a function of the continuous variable
$x=n\psi\in[0,\pi]$,
as in figure \ref{fig:chebplot}. In order to understand better the asymptotic behavior at large $N_{x}$, let us focus on this factor. The function $\big{(}2T_{N_{x}}(a_{n})-2)\big{)}$ is well-behaved, both in modulus and phase, as one can see on figure \ref{fig:chebmodargplot}. The moot point is that the first winding number $n=1$ corresponds to the angle $x=\psi=\f{2\pi}{N_{x}}$ which goes to $x\rightarrow 0$ as $N_{x}$ grows large but not fast enough so as the asymptotics of $\big{(}2T_{N_{x}}(a_{1})-2)\big{)}$ be simply $\big{(}2T_{N_{x}}(0)-2)\big{)}=0$. Indeed the two appearances of the lattice size $N_{x}$ conspires to give a non-trivial asymptotics:
\be
\big{[}2T_{N_{x}}(a_{n})-2)\big{]}^{\f12}\sim e^{(1+i)\sqrt{\f{\pi\lambda N_{x}n}{2}}}
\,,\quad\forall n\ll N_{x}
\,.
\ee
This also gives the behavior for large winding numbers $n\lesssim\lfloor\frac{N_x-1}{2}\rfloor$ since the function $\big{(}2T_{N_{x}}(x)-2)\big{)}$ is (almost) symmetric
\footnote{The symmetry of $\big{(}2T_{N_{x}}(a(x))-2)\big{)}$ depends on the sign of $(-1)^{N_{x}}$. Under the reflection $x\rightarrow \pi-x$, it changes to its complex conjugate for even $N_{x}$ while it further gets an extra minus sign for odd $N_{x}$. 
This means that the modulus  $\big{|}2T_{N_{x}}(a_{n})-2)\big{|}$ is exactly symmetric under $n\leftrightarrow \frac{N_x}{2}-n$ for even $N_{x}$ while it is slightly skewed under the exchange $n\leftrightarrow \frac{N_x-1}{2}-n$ for odd $N_{x}$ due to the $\f12$ shift.
}
under reflections $x\leftrightarrow \pi-x$. This explains the peakedness of the amplitude pre-factor ${\cal A}(n)$ on the two limiting winding numbers  $n=1$ and $n=\lfloor\frac{N_x-1}{2}\rfloor$.


The intuition behind the peakedness at the minimal and maximal winding numbers, $n=1$ and $n=\lfloor\frac{N_x-1}{2}\rfloor$, is that 
these solutions reconstruct locally almost-flat geometries (although one can be visualized as being folded onto itself), and at flat geometries the Hessian degenerates.
This mechanism is analogous to ``Ditt-invariance'' \cite{Dittrich:2012jq,Rovelli:2011fk}.
This peakedness can be used to argue that the slightest (semiclassical) knowledge of the extrinsic curvature, such as the fact that it is non-Planckian as in the maximal $n$ case, collapses the result onto  the desired classical solution at $n=1$.
\footnote{The very same physical argument allows to discard the ``folded'' solutions mentioned above.}

To summarize the role of the amplitude pre-factor ${\cal A}(n)$ is that it selects the first winding number $n=1$, which corresponds to the semi-classical embedding to our toric surface in flat $\R^{3}$ space and allows to reproduce the expected semi-classical partition function \eqref{eq_Z} for 3d quantum gravity as a function of the twist angle $\gamma$.

\section{Outlook\label{sec_outlook}}

In this note, we have proposed a concrete framework to analyze quasi-local holographic dualities in three dimensional quantum gravity.
In such a framework, both the bulk quantum geometry---described by the Ponzano--Regge model---and the boundary theories are readily accessible, and the bulk-boundary correspondence can be readily read from the choice of boundary conditions.
After highlighting the general character of the correspondence and the general questions one hopes to address in its context, we have reviewed the main results of \cite{PRholo1,PRholo2}. 

In this outlook, we wish to discuss more specific questions which arise when studying the holographic setup that is here advocated  for. 

\smallskip

{\bf Boundary phase transitions} The boundary statistical theory might undergo phase transitions for specific choices of the parameters. The question is what this means from the geometrical perspective of the bulk theory. The question is particularly cogent in the case of second order phase transitions. First steps to answer this question have been taken in \cite{Dittrich:2013jxa,Bonzom:2015ova}. There it was shown that---for an infinite superposition of trivalent spin-networks, which turns out to be dual to the Ising model---criticality corresponds to peakedness around geometrical boundary conditions (modulo a global scale).

\smallskip

{\bf Asymptotic infinity} The framework presented here is well-adapted to the study of quasi-local boundaries. Which theory arises when pushing these boundaries to infinity is an independent question. In order to reach infinity, the first guess is that one needs to consider boundaries of infinite circumference. This can a priori be done in two ways: by considering an infinite number of building blocks, possibly Planck-sized (e.g. with $j=1/2$), or by rescaling a given boundary configuration by scaling the relative spins to infinity. The first procedure resembles a continuum limit of the statistical model, provided one at the same time scales down the lattice separation (and rescales the relevant boundary fields). The continuum limit,  however, is usually taken by keeping the total physical size of the system constant.
The second procedure, on the other hand, is related the usual semi-classical limit of spinfoams, which leads to a classical, albeit discrete, theory of gravity. It is possible that the scaling to asymptotic infinity is a mixture of the two procedures above.
%
Given the relationship between continuum limits and second order phase transitions, 
%
identifying the correct notion of asymptotic limit might well shed new light on phase transitions in spinfoam models.
%
%

\smallskip

{\bf Cosmological constant} Introducing a cosmological constant $\Lambda$ seems a natural goal, especially in relation to the (A)dS/CFT correspondence. Euclidean three-dimensional models with a cosmological constant are known and some of their properties have been widely studied \cite{Turaev:1992hq, TaylorWoodward2005}. They correspond to real ($\Lambda<0$) and root-of-unity ($\Lambda>0$) $q$-deformations of the Ponzano--Regge model used in this note and in \cite{PRholo1,PRholo2}. In particular, the model with $\Lambda>0$ coincides with the original Turaev--Viro topological field theory for $U_q(\su(2))$, $q$ root of unity. This model is finite and hence mathematically well defined from the outset, i.e. without the need of gauge fixing (which was implicit in the present treatment, see \cite{PRholo1}). There is therefore little difficulty in generalizing the present treatment to those cases. Difficulties, however, arise when one tries to compare the ensuing geometries to the (A)dS/CFT or even AdS/MERA setups (see also \cite{DittrichDonnellyRiello}). 
The main issue is that the discrete quantum
%
geometry encoded in the Turaev--Viro weights is that of homogeneously curved building blocks.\footnote{Hence zero deficit angles in the corresponding Ponzano--Regge-like model correspond to absence of ``extra'' curvature defects, and thus to a homogeneously curved manifold. See \cite{MizoguchiTada1992,TaylorWoodward2005,Bahr:2009qc,Bonzom:2014bua,Livine:2016vhl} for a treatment in three dimensions. Generalization to four dimensions also exist \cite{Haggard:2015ima,Dittrich2017}.}
While this fact is well adapted to a MERA-like triangulation of the bulk manifold (in the AdS case), it is much less adapted to describe the common conformally-flat boundary at AdS time-like infinity.
Fortunately, this seems to be more a nuisance than a fundamental problem.
Matching the boundary theories obtained in AdS/CFT and in the present setup seems to be a more cogent problem, which might however be inextricable from the previous problem on how to encode asymptotic infinity in the present setup.

\smallskip

{\bf BMS$_\mathbf{3}$} To conclude, let us mention one more issue. Although the calculation of the partition function nicely matches previous calculations and in particular the formal expression of BMS$_3$ characters of Oblak \cite{Oblak:2015sea}, it is still unclear whether and in which sense BMS$_3$ emerges (in the continuum limit) as a symmetry of the boundary theory discussed above. Of course, this is a crucial question to answer in order to claim full understanding of the setup reviewed in the previous section.

\acknowledgements{
\vspace{-.5em}
This work is supported by Perimeter Institute for Theoretical Physics. Research at Perimeter Institute is supported by the Government of Canada through Industry Canada and by the Province of Ontario through the Ministry of Research and Innovation.

Plots were performed using Mathematica 10.
}

\bibliographystyle{bibstyle_aldo}
\bibliography{PRholo_biblio}

\end{document}